\documentclass[onecolumn,epsfig,pre]{revtex4}
\usepackage{graphicx}
\usepackage{amssymb}
\usepackage{amsmath}
\usepackage{hyperref}
\usepackage{latexsym}

\def\mdiv{{\mbox{div}}}
\def\mcurl{{\mbox{curl}}}
\draft

\begin{document}
\title{Chiral tilt texture domains in two dimensions}
\author{Sarasij R. C.$^{1,2}$ and Madan Rao$^{1,2}$}
\affiliation{
$^1$Raman Research Institute, C.V. Raman Avenue, Bangalore 560080,
India\\
$^2$National Centre for Biological Sciences, Tata Institute of Fundamental Research-GKVK Campus, Bangalore 560065, India}
\begin{abstract} We study the shape and texture of finite domains comprising chiral/achiral molecules carrying a tilt field embedded in a 2-dimensional surface. Using a combination of simulations and {\it exact} variational calculations, we determine the equilibrium (mean field) phase diagram in the achiral and chiral cases. We find a variety of novel shapes and textures including a {\it spontaneously broken chiral texture} when
the molecules are achiral. We show that chiral tilt-domains nucleating in 
a region of two-phase coexistence, {\it repel} each other, thereby preventing coalescence and further growth.
Our work  has implications for tilt texture domains in phospholipid monolayers and giant unilamellar vesicles (GUVs), nucleating domains  of Sm-C$^*$ in Sm-A films, and chiral emulsions in Sm-A films.
\end{abstract}
\maketitle

\section{Introduction} 
Shapes and textures of finite-size domains embedded in 
two or three dimensional substrates are governed by an interplay between elasticity, defects and finite size. Such self-assembled patterning gives rise to unique rheological and optical properties with potential engineering applications as in the production of templates, micro-lithography, liquid-crystal displays and optical switches \cite{choi}. The purpose of this paper is to provide a detailed analytical and numerical study of the shapes and textures of finite-size domains in 2-dimensions when the constituent molecules (or molecular aggregates) are both chiral and possess a tilt \cite{molorient,tilt}. This has implications for a wide variety of soft materials such as (i) domains in freely suspended phospholipid monolayers in an air-water interface \cite{monolayer,LANG,knobler,selinger}, (ii) domains in tense lipid bilayer vesicles \cite{vesicle,webb,raghu} or freely suspended bilayers in an aqueous solvent \cite{bilayer}, (iii) nucleation of Sm-C$^*$ domains in freely suspended Sm-A films at the A-C$^*$ transition \cite{salanger}, (iv) droplets of a suspension of ferromagnetic particles in a liquid film \cite{langergold} and (v) emulsion of chiral nematogens in a liquid film. A variety of techniques such as fluorescence confocal polarization microscopy, atomic force microscopy, Brewster-angle microscopy and Bitter microscopy (in the case of ferrofluids) have been used to detect domain shapes and textures on 2-dimensional substrates. Moreover the tilt phase of chiral molecules with an electric dipole moment is ferroelectric \cite{meyer,degennes}, allowing for the possibility that external electric fields may be used to bias the textures \cite{electric}.

Over the years there has been a substantial amount of experimental and theoretical work in this area. However, most theoretical studies have been restricted to the case when the constituent molecules are achiral. In this paper we systematically study the effects of chirality on the shape and texture of tilt domains. The competition between bulk orientational order, chirality and anisotropic line tension from the boundary gives rise to a variety of shapes and textures. Even in the achiral limit we point out several new phases missing in earlier analyses, such as the {\it annular phases}, which we establish using an `exact' variational approach (whose meaning will be made clear later on). Unlike previous studies, our variational calculation allows the {\it topology} of the domain to vary, {\it i.e.}, the domain need not be simply connected.
The method of analysis followed in this paper takes off from and extends the elegant treatment of Ref.\,\cite{pettey}.

In the next section, we discuss the various contributions to the domain energy containing chiral molecules with an orientational field. We will 
arrive at an effective continuum hamiltonian (Sect.\,2.1) describing the shape and texture of this finite, two-dimensional domain.  We will next discuss (Sect.\,3) the optimum shape and orientational
ordering within the finite patch of fixed area. 
The main results of this paper are --- (i) We obtain an `exact' phase diagram in the achiral limit when the boundary of the domain is circular (Sect.\,3.1), with a  variety of optimal textures such as {\it virtual boojum}, defect, achiral annular, and a {\it spontaneously broken} chiral annular phase. (ii) We study smooth perturbations of the boundary, determine the regime of instability of the circular domain (Sect.\,3.2), and the nature of the resulting equilibrium phase. We find that the annular phases are no longer minimum energy configurations. Typical domain shapes are thin and elongated, best described as being rectangular with circular caps. (iii) Turning on bulk chirality gives rise to a spiral defect phase and a novel {\it chiral tweed} phase (Sect.\,3.3). (iv) Chirality may even  induce a large enough domain to split into multiple domains; we determine the conditions under which such multiple domains obtain (Sect.\,4). This suggests that tilt-domains, nucleated following a quench across a phase-boundary in a multicomponent system, would repel each other, preventing coalescence and subsequent growth.
Finally we discuss how an electric field might couple to the tilt when the molecules are chiral; this may be used to switch between degenerate textures (Sect.\,6). In the following sections we will see  how these results are obtained.

\section{Construction of an effective hamiltonian} 

The question we address is : what is the equilibrium texture of a collection of chiral molecules described by an orientational field, uniformly spread over  a finite domain of fixed area $A$ embedded in a 2-dimensional (2d) flat substrate ? We show that the interplay between orientational ordering and domain shape gives rise to a variety of novel phases and shape transitions. We will first discuss the nature of the order parameter and the dominant contributions to the energy of the domain.

The orientation of a rigid molecule may be described by a polar vector which takes values in S$^2$ (Heisenberg spin) or a director which takes values in the projective space RP$^2 \equiv$\,\,S${^2}/$Z$_2$ \cite{condmat}. However energetic considerations, e.g., in the case of membranes, a combination of van der Waals and hydrophobic shielding, may constrain (a) the centre of mass of the molecules to lie on a 2d, flat surface and (b) the projection of the long-axis of the molecule onto the 2d plane to have a fixed magnitude (or small deviations from a fixed value). Thus
owing to strong uniaxial anisotropy, the low energy sector may be described by a 2d  polar vector ${\bf m}$ (since the 2d surface carries a unique local outward normal) which takes values in S$^1$, an XY spin.
The centre-of-mass density $\rho(x,y)$ of the molecules is assumed to be uniformly smeared over the patch of area $A$. 
 In addition, individual molecules may have a permanent dipole or quadrupole moment --- this is especially true in the case of zwitterionic lipids (electric dipole moment) or ferromagnetic particles (magnetic dipole moment). The molecules outside the finite patch are assumed to be either in the isotropic or the Sm-A (i.e., no tilt) or the liquid-disordered ($L_d$) phase.

The tilt molecules interact with each other, and with the molecules outside the patch, both sterically (purely repulsive) and via short range (e.g., van der Waals) attractive interactions. Both these effects contribute to chiral interactions; the former via the Straley picture of interlocking screws \cite{straley},  the latter via a generalisation of the Van der Waals dispersion to chiral molecules \cite{harris}. In the continuum limit, these short-range interactions can be written as the usual
Frank energy, modified to include the effects of chirality. In addition to these short-range interactions there could be long-range dipole-dipole (or higher multipole) interactions between the tilt molecules carrying a permanent dipole moment, which may also contribute to chiral interactions.

In this paper, we ignore the contribution coming from dipolar interactions, which would be justified
if they are demonstrably smaller than the Frank terms (see Sect.\,5).
For instance, in the case of phospholipid domains, the Frank elastic constants are of order $10 k_{B}\,T_{room}$, while the dipolar energy per phospholipid is about $10^{-22}$\,J, and hence may be ignored. In situations where dipolar energy scales are comparable, it is likely to have a major effect on domain shapes and textures, such as, for instance, the formation of string-like aggregates \cite{yethiraj}.

In our present treatment we assume the substrate to be flat over the scale of the patch; thermal undulations of the substrate (liquid film or membrane) controlled by surface tension or bending rigidity are assumed to be negligible over this scale. Elsewhere we have discussed the effect of chiral tilt textures on a {\it flexible membrane}, leading to a novel chirality induced budding and tubulation \cite{budding,bpj}.

\subsection{Tilt texture hamiltonian}
Let us assume that the molecules, now described by a 2d orientational field ${\bf m}$, are smeared with a fixed, uniform density over a patch (domain) of area $A$ embedded in a 2d surface, the $xy$ plane.
The rest of the surface surrounding the patch is a structureless fluid consisting of molecules different from the constituents of the patch (or else in a different phase) or a solid substrate. We assume that the patch constituents have come together as a result of micro phase segregation. The domain energy can be written in terms of bulk distortions of ${\bf m}$ and interfacial (perimeter) distortions, including an orientation-interface coupling. 

In our low energy description, the magnitude of  ${\bf m}$ is held fixed and normalised to unity (hard spin model). The resulting domain energy only
contains {\it phase} distortions of ${\bf 
m} \equiv (\cos \phi, \sin \phi)$. Such a {\it phase-only} theory is formally ultraviolet divergent, and needs to
be made finite by a microscopic cut-off length of molecular
dimensions. The regime of validity of such phase-only theories is over scales larger than this
cut-off length. At these scales, the hard-spin model is 
equivalent to a soft-spin model, where the magnitude of ${\bf m}$ is constrained by means of a potential $V({\bf m}) = u\, ({\bf m}\cdot{\bf m} - 1)^2$  to deviate only slightly from its preferred value, for $u$ large and postive.

The form of the bulk distortion energy of ${\bf m}$ can be constructed from symmetry arguments. Since tilt is a spatial vector, tilt distortions may be described by a hamiltonian invariant under 2d spatial rotations $O(2)$. 
To lowest order, this leads to the usual Frank-energy in 2d, with two independent contributions --- a {\it splay} distortion $(\nabla \cdot {\bf m})^2$ and a {\it bend} distortion $(\nabla \times {\bf m})^2$, where $\nabla \equiv (\partial_x , \partial_y)$. Recall however that the molecules are chiral, and so on general symmetry grounds, we should allow for terms which are invariant under improper rotations alone; thus to lowest order this gives rise to a term of the form $(\nabla \cdot {\bf m})\,
(\nabla \times {\bf m})$. Note that for a 2d vector field ${\bf m}$,
$\nabla \times {\bf m}$ is a pseudo-scalar, not a vector. Viewed as a vector in $R^{3}$, the curl\,${\vec m}$ is a pseudo-vector pointing along ${\vec z}$ (we will consistently represent 2d vectors by boldface and 3d vectors by an overhead arrow).

The coupling between the tilt and interface gives rise to boundary terms --- if ${\bf n}$ is the local unit normal to the boundary aiming into the domain,
then there are two possible boundary terms to linear order in ${\bf m}$. These are (i) an isotropic line tension proportional to the perimeter and (ii) anchoring terms of the form $({\bf m}\cdot {\bf n})$ and $({\bf m} \times {\bf n})$, which contribute to an effective anisotropic line tension. One source for this boundary term is that the interface between the tilt and non-tilt components (in the case of chiral emulsions this coincides with the tilt-solvent interface) prefers a specific alignment of the tilt with
the local normal. Alternatively such boundary contributions may arise from the presence of a bulk
spontaneous bend $\nabla \times {\bf m}$ and splay $\nabla \cdot {\bf m}$, since
\begin{eqnarray}
\int_{A} {\mbox{div}}\, \vec{m} = 
\oint_{C} \vec{m} \cdot \vec{n} \\
\int_{A} {\mbox{curl}}\, \vec{m} = 
\oint_{C} \vec{m} \times \vec{n} \, .
\end{eqnarray}
The spontaneous bend is a chiral contribution;
spontaneous splay could arise from specific substrate-tilt coupling or from steric forces arising from large head-to-tail ratio of the constituent molecules.

Thus our low energy effective hamiltonian for 
the chiral tilt domain in the {\it strong segregation limit}, is given, to quadratic order in the fields, by 
$E=E_B+E_C$, where the bulk energy 
\begin{equation} 
E_{B} = \int_{A} \left[ \frac{k_{1}}{2} 
(\nabla\cdot{\bf{m}})^{2}+\frac{k_{2}}{2}(\nabla\times{\bf{m}})^{2} 
+ k_{c}(\nabla \cdot{\bf{m}})(\nabla \times {\bf{m}}) \right]\, d^2x\, , 
\label{eq:bulk} 
\end{equation} 
and the interfacial energy, 
\begin{equation} 
E_{C} = \oint_{C} dl \left(\sigma_{0}+\sigma_{1}(\bf{m} \cdot \bf{n})+ 
\sigma_{2}(\bf{m} \times \bf{n})\right) \, .
\label{eq:interface} 
\end{equation} 

When the coefficients $k_1$ and $k_2$ are comparable, the Frank terms resist any deviation of ${\bf m}$ from uniformity and hence the ${\bf m}$ field would everywhere point in a specific direction. The anisotropic
boundary terms however prefer to align 
${\bf m}$ at the boundary along (or orthogonal to) the local normal. This competition gives rise to a variety of nontrivial textures. The boundary terms may be interpreted as providing an effective anisotropic line tension of the form
$\sigma_{0}+\sigma_{1}(\bf{m} \cdot \bf{n})+ 
\sigma_{2}(\bf{m} \times \bf{n})$. A negative value of this effective line tension would lead to an instability of the circular domain boundary \cite{pettey}.
Nontrivial textures also obtain when the Frank coeffcients $k_1$ and $k_2$ are appreciably different \cite{fournier}. We will see that chirality introduces a new twist to the texture phase diagram \cite{budding,bpj}.

In this strong segregation limit, molecules are not allowed to exchange across the domain. This imposes a constraint on the allowable configurations explored by the molecules. Thus for fixed ``bath'' conditions (state of the molecules outside the patch), the configuration of molecules inside the patch can reach equilibrium. The area of the patch and the density of molecules in the patch are fixed over this time scale. In the case of nucleation of the chiral tilt domains (e.g., Sm-C$^*$ domains in a freely suspended Sm-A film or tilt domains of chiral lipids in a mono/bilayer), slow domain growth justifies the use of quasi-equilibrium ideas. For a dilute emulsion, the area of the domain is fixed at time scales smaller than domain coalescence times; thus the shape and texture of the domain will be such as to minimise the free-energy subject to the constraint of constant area.

Our aim then is to find that optimal conformation, in general a difficult calculational task. We will see that given the energy scales in the problem (Sect.\,5), we may ignore the effects of thermal fluctuations on the shape and texture of the domain.
Thus, the Frank constants $k_{1}, k_{2}$ are $\sim 10 k_{B}\,T_{room}$ while the 
interfacial energy of a domain of radius $10\,\mu$m is an order larger. Hence a  mean field minimisation of the above free-energy functional will suffice.

\section{Mean field phase diagram} 

We will perform this constraint minimisation of the free energy using a variational approach, where we vary both the texture and the domain shape while keeping the area constant. Our constrained variational ansatz also accounts for
the possibility that the domain may not be simply connected, however
we will not allow the domain to break up (domain splitting will be treated in Sect.\,4). Our variational guesses are supported by
computer simulations, i.e., monte carlo simulations with simulated annealing to avoid 
getting 
stuck in local minima. It is in this sense that our variational calculation for the mean field phase diagram is `exact'  (made more explicit in the next subsection).
We have also explicitly checked that the effect of including higher order (symmetry allowed) terms in the hamiltonian is small and does not affect the phase diagram. 

We  henceforth set $k_{1} = 1$ as our unit of energy 
and $R=1$, associated with the domain size $R = 
\sqrt{A/\pi}$, as our unit of length. Some experimental situations are more conveniently analysed by fixing $\sigma_0=1$ as the unit of length 
and determining the phase structure in the $(R,\sigma)$ plane. We discuss both these ensembles.

\subsection{Achiral domain}

We first turn off the effects of bulk chirality ($k_c = 0$), but retain the anisotropic boundary term. In most of our work we take $k_{2} = k_{1} = K$ (the one-coupling-constant approximation); towards the end we comment on the distinct features when $k_1 \neq k_2$.

In terms of the phase angle $\phi$, where ${\bf m} \equiv (\cos \phi, \sin \phi)$, the free-energy functional simplifies to
\begin{equation}
E  =  \frac{K}{2} \int_{A}
(\nabla \phi)^{2} + \oint dl \left(\sigma_{0}+\sigma_{1} \cos(\psi - \phi) + 
\sigma_{2} \sin(\psi - \phi)\right) 
\label{eq:phase}
\end{equation}
where we have written the normal ${\bf n} \equiv (\cos \psi, \sin \psi)$ in polar coordinates.
Variation of the above hamiltonian gives the  {\it bulk} Euler-Lagrange
equation, $\nabla^2 \phi=0$ (Laplace) and nontrivial boundary equations which have to be simultaneously satisfied.
Instead of asking for solutions of the combined bulk+boundary equations, we determine the global minimum free-energy configuration variationally --- this has the advantage of obtaining {\it boundary minima}, minimum energy configurations that may not be solutions of the Euler-Lagrange equations.

Without loss of generality we may work with a simplified free-energy functional --- as argued by previous authors \cite{salanger,pettey}, a global rotation of ${\bf m}$ by
$\phi^{\prime} = \phi - \tan^{-1} (\sigma_2/\sigma_1) - \pi/2$
without rotating the boundary contour transforms the free-energy functional  to a new functional
\begin{equation}
E  =  \frac{K}{2} \int_{A}
(\nabla \phi)^{2} + \oint dl \left(\sigma_{0}+\sigma \cos(\phi-\psi) \right)
\label{eq:gauge}
\end{equation}
where $\sigma = {\sqrt{\sigma_1^2 + \sigma_2^2}}$ is the single redefined anisotropic boundary coefficient. 
Configurations which minimise the free energy (\ref{eq:phase}) with parameters $(K,\sigma_0,\sigma_1,\sigma_2)$ can be obtained from configurations which minimise the free energy (\ref{eq:gauge}) with parameters  $(K,\sigma_0,\sigma)$ via the global transformation defined above. Thus in this equal-constants, achiral regime, it suffices to work out the phase diagram in the $(\sigma_0, \sigma)$ plane using the free energy (\ref{eq:gauge}), the phase diagram for arbitrary values of $\sigma_1, \, \sigma_2$ can then be reconstructed. This simplification does not occur when the bulk chiral term is present. 
\begin{figure}
\begin{center}
\includegraphics[width=4in]{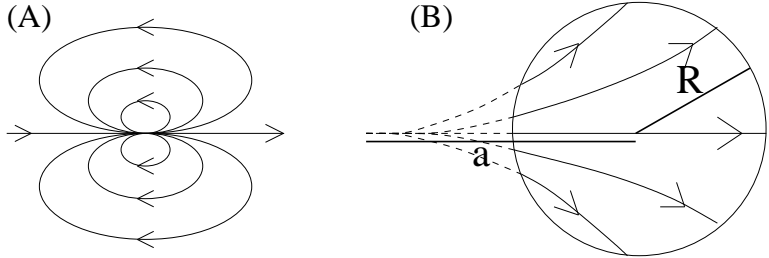}
\caption{Effect of anisotropic line tension on the texture of an achiral circular domain : (A) a boojum of charge 2, (B) texture of a domain of radius $R$ whose centre is at a distance $a$ from the core of the boojum. The arrows give the local direction of ${\bf m}$.}
\label{texture}
\end{center}
\end{figure}
Note that by making a global rotation, it is possible to entirely elimate the 
contribution from the boundary chiral term $\nabla \times {\bf m}$. This implies that the domain shapes will always be achiral \cite{pettey}.

Now set $R=1$, as our unit of length.
We will first discuss the limit of large $\sigma_0$,  when the domains are forced to be circular (we will discuss the limit of stability of a circular domain in  the next section). The extreme limit, $\sigma_0 \to \infty$, might be experimentally arranged by smearing a thin film
of Sm-C$^*$ over a hole made in a solid substrate.
Clearly in the absence of any anisotropic line tension, $\sigma =0$, the lines of ${\bf m}$ describing the optimal texture
are a (parallel) set of straight lines pointing in {\it any} direction. This configuration is invariant under arbitrary translations and rotations of the boundary circle with respect to the lines of ${\bf m}$.
An infinitesimal (positive) $\sigma$ forces the lines of ${\bf m}$ to curve slightly (at the cost of Frank energy) to meet the circular domain boundary at a desired angle (Fig.\,1). From the axial symmetry of the solution, it is clear that this configuration is associated with a $+2$ virtual defect, situated {\it outside} the domain, called a {\it virtual boojum} \cite{salanger}. 
Note that this configuration is infinitely degenerate. To calculate the energy we need to parametrize the 
virtual boojum configuration, where the 
centre of the circular domain is a distance, $a$, away from the core of a 
virtual defect of strength $N$. Placing the origin of polar coordinates 
$(r, \theta)$ at the core of the boojum, we can describe the texture ${\bf 
m} \equiv (\cos \phi, \sin \phi)$ by the equation 
\begin{equation} 
\phi = N \theta + c_{1} r 
\sin \theta + c_{2} r^{2} \sin 2\theta + c_{3} r^{3} \sin 3\theta + \ldots
\label{eq:texture} 
\end{equation} 
The first term is the singular part representing the boojum, subsequent terms represent the most general smooth solution of Laplace's equation with the desired symmetry.
The energy $E$ is minimised 
with respect to the variational parameters $(N, a, \{c_{n}\})$ for fixed 
$(\sigma_0, \sigma)$. Consistent with our symmetry arguments, we explicitly find that the optimal value of $N$ is exactly $2$ and
independent of $a$. 
\begin{figure}
\begin{center}
\includegraphics[width=2in]{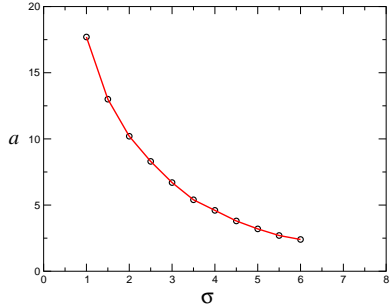}
\caption{Variation of $a$, the distance of the virtual boojum from the centre of the domain, as function of
the anisotropic line tension $\sigma$ : note that the penetration of the boojum core into the domain is pre-empted by the annular phase. We have taken $\sigma_0=1$ for convenience.}
\label{parameter-a}
\end{center}
\end{figure}
In principle ofcourse we can take as many variational parameters $\{c_n\}$ as we desire, however
in drawing the phase diagram we have parametrised the 
texture by $a$ and $c_{1}$ alone; inclusion of higher order $c_{n}$
 lowers the energy by $1\%$ at most. The form of the variational energy for $\phi = 2 \theta$ is given by \cite{pettey}
 \begin{equation}
 E(a) = - 2 \pi K \ln \left[1-\left(\frac{R}{a}\right)^2\right] - 2 \pi 
 \frac{\sigma R^2}{a} + 2 \pi \sigma_0 R
 \end{equation}
 which is to be minimised with respect to $a$. 
The minimisation with respect to the whole set of parameters $(a, \{c_n\})$, however, is best carried out numerically. Upon variation, we find that
an increase in the anisotropic line tension $\sigma$ leads to an increase in $c_1$, pushing the core of the virtual defect towards the domain centre. Experimental characterisation of this texture in the context of Langmuir monolayers at the air-water interface \cite{LANG,knobler} agrees with our analysis above.
This movement of the core continues till $a=1.7$, beyond which there is a discontinuous transition to
the annular phase (Fig.\,2). Note that the boojum is always virtual, the core never penetrates the domain, since it is pre-empted by the annular phase.
Reference \cite{pettey} misses this feature, since their variational shapes do not include such multiply connected topologies.

The {\it Achiral Annular} phase can be parametrized by an inner and 
outer radii $r_1$ and $r_2$ respectively. At the outer rim of the annulus,
${\bf m}$ is directed radially outward, while at the inner boundary, ${\bf 
m}$ is inclined at an angle $\alpha$ to the local normal $\bf{n}$. With 
the origin of polar coordinates at the centre of the annulus, the texture 
may be described by 
\begin{equation}
\phi = \theta - \alpha \frac{r_2-r}{r_2-r_1}\, , 
\end{equation}
where 
$r_{1}, \alpha$ are variational parameters ($r_2$ may be obtained from the constant area constraint). With this parametrisation, the energy of this configuration
is given by,
\begin{equation}
E(r_{1}, \alpha) = \pi K \left[\frac{\alpha^2}{2} \left(\frac{r_2+r_1}{r_2-r_1}\right)
+ \ln \frac{r_2}{r_1}\right] - 2 \pi \sigma \left(r_2-r_1\cos \alpha\right)
+ 2 \pi \sigma_0 \left(r_1 + r_2\right)
\label{eq:achiannen}
\end{equation}
where $r_2^2 - r_1^2 = R^2$.
The inner radius $r_1$ 
monotonically decreases as we increase $\sigma$ (Fig.\,3) such that $\bf{m}$ always 
points radially outward ($\alpha = 0$) at every point of the domain. With increasing $\sigma$, we 
cross a first order phase boundary (Fig.\,4) into a new 
annular phase with a chiral texture.
\begin{figure}
\begin{center}
\includegraphics[width=2in]{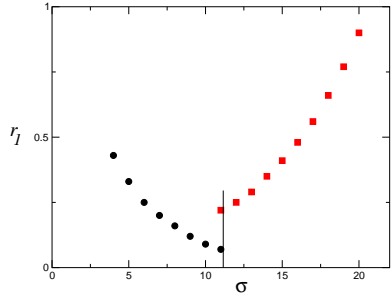}
\caption{The inner radius $r_1$ of the annular domain 
decreases (increases) with $\sigma$ in the achiral (chiral) phases (represented by circles (squares)). We have taken
$\sigma_0=1$ for convenience.}
\label{parameter-r1}
\end{center}
\end{figure}

The {\it Chiral Annular} phase obtains when the angle $\alpha$ jumps to a value greater than $\pi/2$ at 
both the inner and outer boundaries. The bulk texture is chiral and is a true symmetry broken configuration (recall that there is no bulk chiral interaction) with the pseudoscalar order 
parameter 
\begin{equation}
C = \frac{1}{R}\int d^{2}\,{x}\, (\nabla \times 
\bf{m}) \neq 0 \, .
\label{chiop}
\end{equation}
This chiral phase is doubly degenerate and can spontaneously acquire 
either sign; the anisotropic tension $\sigma$ behaves as a {\it surface} field conjugate to this Ising-like chiral order parameter. 

Note that the `core' of these annuli must consist of molecules without tilt; our assumption in minimising the energy has been that the two species of molecules are free to diffuse across the domain of fixed area $A$.
If for some reason the diffusion of the non-tilt molecules from outside the patch to the `core' of the annulus is hindered, then the appearance of such a phase would be kinetically blocked. In this case, the virtual boojum will penetrate the domain and would lead ultimately to the Hedgehog configuration (as in \cite{pettey}).

The  {\it Hedgehog} phase has a defect at 
the domain centre and a texture described by $\phi = \theta$ (when the origin is at the domain centre). The energy 
includes a core contribution $\epsilon_{c}$ of the defect of {\it microscopic} size 
$r_{c}$ ($r_c \ll R$),
\begin{equation}
E_{hedge} = \pi K \ln \frac{R}{r_c} - 2 \pi R \left(\sigma - \sigma_0\right) + \epsilon_{c}
\label{eq:hogen}
\end{equation}

The phase diagram in the $(R, \sigma)$ plane is shown in Fig.\,4. The transitions between these phases are discontinuous, as indicated by the change in slope of the energy branches as a function of $\sigma$. 
As discussed earlier, in some experimental situations it is more useful to study the phase diagram in the 
$(R,\sigma)$ plane; thus we set $\sigma_0 = 1$ to obtain a unit of length. Assuming that the domain is still
circular, the phase diagram shows the discontinuous
transitions which
weaken (i.e., the jump in the appropriate order parameter decreases) as the domain size $R$ shrinks. 
\begin{figure}
\begin{center}
\includegraphics[width=2in]{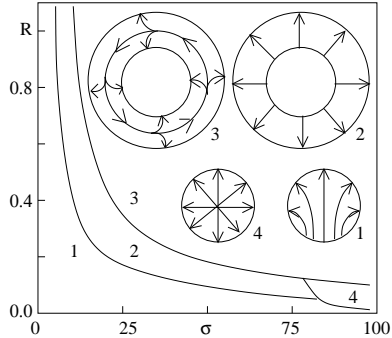}
\caption{Phases of the tilt texture domain with circular 
periphery\,: (1) Virtual Boojum (2) Achiral Annular (3) Chiral Annular 
(4) Hedgehog with 
$\epsilon_{c} = 0$. Arrows indicate direction of {\bf m} field. }
\label{fig1}
\end{center}
\end{figure}

This analysis suffices to reconstruct the entire phase diagram in the $(\sigma_1, \sigma_2)$ plane by global rotation of optimal textures obtained above. For instance, the virtual boojum texture reported in \cite{pettey} can be obtained from a global rotation of the texture in 
Fig.\,1. Likewise the {\it Vortex} phase can be obtained by globally rotating the hedgehog texture by an angle $\pi/2$. The vortex texture spontaneously breaks chiral symmetry. Other nontrivial phases may be obtained by global rotations of the texture in the annular phases.

 Finally we mention in passing, the case when $k_1 \neq k_2$, which has been the subject of study in \cite{fournier}. The difference in the splay and bend energies implies that the hedgehog and vortex defects are no longer degenerate. If $\Delta K = (k_1 - k_2)/k_1$, then as $\Delta K$ goes from large negative to large positive values, the texture goes from a hedgehog to a vortex via a spiral defect phase.

\subsection{Noncircular domains}
For smaller values of $\sigma_0/\sigma_1$, the domain is no longer circular; 
as shown by \cite{pettey,bruinsma}, deviations from circularity arise when the effective line tension $\sigma_{0}+\sigma_{1}(\bf{m} \cdot \bf{n})+ 
\sigma_{2}(\bf{m} \times \bf{n}) < 0$. We summarize their results,  based on linear stability analysis.

Parametrize the boundary by smooth 
deformations of a circle, $r(\theta)$ where we have temporarily shifted the origin from the core of the
boojum by a distance $a$ to the centre of the domain, i.e.,  the centre 
of the circle of radius $r_0$ and $0 < \theta < 2 \pi$ is the angle to the 
polar axis, 
\begin{equation}
r(\theta) = r_0 \,\left( 1 + 
\sum_{n=1}^{\infty} \alpha_{n} \cos\, n\theta \right).
\end{equation} 
Such a parametrization only includes shapes with no overhangs, i.e., the coordinate $r(\theta)$ is a single-valued function of  $\theta$. Further the parametrization is smooth and so does not include shapes with cusps, as in \cite{bruinsma,knobler}. Since the area of the
domain has to be $A$, we must have
\begin{equation}
\pi r_{0}^{2} \left( 1 + \frac{1}{2} \sum_{n=1}^{\infty} \alpha_{n}^{2} \right) = A = \pi r^{2}
\end{equation}
A perturbative analysis \cite{pettey} involves perturbing the domain boundary about the circle {\it and} the texture about the optimal texture (virtual boojum, annuli or hedgehog). Such an analysis, carried out in \cite{pettey} for the virtual boojum, indicates that the 
circle is unstable to the $n=2$ mode when $\sigma > \sigma_0$, below a critical value of $K/\sigma R$.

We complement the linear stability analysis by a variational calculation. The reason we do this is because (a) linear stability analysis typically underestimates phase boundaries and (b) linear stability analysis does not give the (new) stable configuration in the regime of instability of the circle. To address these issues, we parametrise both the boundary and texture and determine the lowest energy configurations of the texture and the domain shape using a variational scheme. 

Again we set $K = \sigma_{0} = 1$ to fix the units of length and energy. We now minimise the energy as a function of variational parameters \{$N$, $a$,
$\{c_{n}\}$, $\{\alpha_{n}\}$\} for a given value of $R$ and $\sigma$.
The optimal value of $N$ is, as before, $N=2$. We also find that including
terms in the texture with coefficients $\{c_{n>1}\}$,
reduces the minimum value of $E$ by at most $1\%$. 
So we parametrise
the texture in the simpler form
\begin{equation}
\phi = 2 \theta + c_{1} y
\end{equation}
where $y$ is the distance of any point from the axis of the domain.
As $\sigma$ is increased from $0$, the core of the boojum approaches
the centre of the domain but the domain remains almost circular. 
\begin{figure}
\begin{center}
\includegraphics[width=4in]{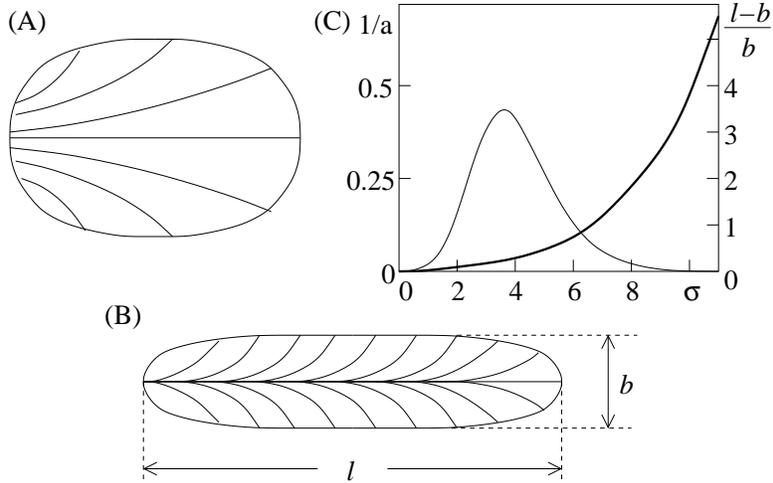}
\caption{Stretching action of strong anisotropic line tension --
domain conformation for (A) $\sigma=6.5$, (B) $\sigma = 9.5$. (C) variation of $1/a$ (thin line) and the
prolateness $(l-b)/b$ with $\sigma$ for a domain of
effective radius $R = 1$. }
\label{prolate}
\end{center}
\end{figure}

As $\sigma$ is raised beyond a threshold $\sigma^*$, the core {\it begins to recede from
the centre} and the domain bulges out at the equator (Fig.\,5) and flattens near the poles. Indeed we find that the 
texture throughout the $(R, \sigma)$ plane is either a virtual boojum or a hedgehog defect.
For large values
of $\sigma$, but still in the boojum phase, it is a good approximation to set $\phi = c_{1} y$ and
take the domain to be a rectangle bracketted at the equator
by two semicircular caps (Fig.\,5). 

Such rectangular domains have been observed \cite{sanat} on giant unilamellar vesicles (GUV's) consisting of a binary lipid mixture, a saturated component (e.g., DPPC) and an unsaturated component (e.g., DOPC), at temperatures below the gel transition of the saturated lipid (e.g., 42$^\circ$\,C for DPPC). Following a temperature quench, domains of the saturated lipid, observed using fluorescence microscopy, were found to nucleate in the background of the unsaturated lipid and grow as rectangular domains.
We believe that the tilt of DPPC in the gel phase gives rise to an anisotropic line tension, and can account for the shape of these domains, consistent with our analysis.

\subsection{Chiral domain} 

We now turn to a description of the phases with nonzero bulk chirality 
$k_{c}$. In this case, both the texture and the shape of the boundary
may assume chiral shapes. For the moment however, let us assume that the boundary of the patch is circular. 
As before we set $k_{1} = k_{2} = 1$ and $\sigma_{0} = 1$. Further without loss of generality we take $k_c >0$. To highlight  the effects of bulk chirality, we have set the anisotropic line tension $\sigma = 0$. It will be clear that the effects of bulk chirality dominate the nature of the texture.

Let us rewrite the Frank expression as
\begin{equation}
E = 2 \pi R + \int_{A} \frac{1}{2} \left({\mdiv}\,{\bf{m}} + {\mcurl}\,{\bf{m}}\right)^{2}
+ (k_{c} - 1) ({\mdiv}\,{\bf{m}}) ({\mcurl}\,{\bf{m}})
\label{chiral}
\end{equation}
It is clear, and we have checked this explicitly, that for low enough bulk chirality, $\vert k_{c}\vert < 1$, the 
{\it Uniform} phase, with a circular domain  and uniform $\bf{m}$, is the lowest energy state. Indeed had we 
included the anisotropic line tension, we would have seen that the optimal achiral domains described in Fig.\,4 would be the lowest energy configurations for small enough $\vert k_c \vert$.

As seen  in (\ref{chiral}), increasing the magnitude of $k_c$, makes the optimal texture more wound up :  the texture prefers to have a very high curl
and a divergence equal and opposite to the curl. The optimal texture is 
neither a pure divergence nor purely rotational, but an (Archimedes) spiral
with an angle of opening $\alpha = \pi/4$ with respect to the local radial (Fig.\,6). 

 In polar coordinates, with the origin at
the centre of the circular domain, the {\it spiral hedgehog} is described by an ${\bf m}$ with constant radial and tangential components everywhere,
$m_{r} = \cos \alpha$ and $m_{\theta} = \sin \alpha$ ($m_{r}^{2} + m_{\phi}^{2} = 1$). Using 
\begin{eqnarray}
{\mdiv}\,{\bf{m}} = \frac{m_{r}}{r}
\\
{\mcurl}\,{\bf{m}} = \frac{m_{\phi}}{r}
\end{eqnarray}
we may compute the energy of this spiral configuration,
\begin{equation} 
E = 2\pi (r_{c}+R) - \frac{1}{2} \left(k_{c}\sin 
2\alpha - 1\right) \ln \frac{R}{r_{c}} + \epsilon_{c} 
\label{eq:energy1} 
\end{equation} 
where $r_{c}$ is
the radius and $\epsilon_{c}$
is the energy of the defect core. 
The optimum value of $\alpha$ is $\pi 
/4$ when $\vert k_{c}\vert > 1$, making $m_{r} = 1/\sqrt{2}$, $m_{\phi}
= - 1/\sqrt{2}$. The spiral hedgehog phase clearly gives a nonzero value 
for the chiral order parameter $C$ (\ref{chiop}). 

We see that the effect of the $k_c$-term is felt most strongly at the core of the spiral defect, falling off inversely as the square of the distance from the core. 
Can we find a texture whose
chiral strength is large not just at one point but
over the entire domain ? If so, this would surely be a candidate for the optimal
texture.
\begin{figure}
\begin{center}
\includegraphics[width=3in]{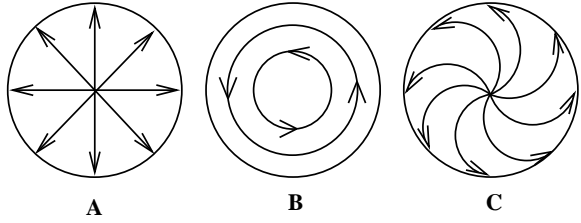}
\caption{(A) Pure divergence (B) Pure rotational (C) Spiral defect. }
\label{spiral}
\end{center}
\end{figure}

To answer this question, we have carried out a
Monte Carlo simulation of $3055$ particles carrying O(2) spins on a 
triangular lattice with a hamiltonian obtained by discretising $E$. In order to search for the lowest energy configuration,
we had to resort to simulated annealing 
from $(k_B\,T)^{-1} = 0.1 \to 300$ starting from a variety of initial 
conditions.  The energy functional is augmented by a
higher order term 
$\gamma\,(({\mdiv}\,{\bf{m}})^{4}+({\mcurl}\,{\bf{m}})^{4})$ to provide a cutoff to the spatial variations of ${\bf m}$.
Fig.\,7 shows the optimal texture obtained for 
$\sigma = 0$ and $k_{c} = 12.75$ --- the texture is best described as a {\it chiral tweed} with a stripe size $l^*=0.01$.
\begin{figure}
\begin{center}
\includegraphics[width=3in]{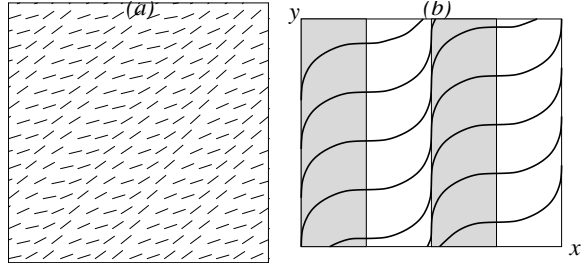}
\caption{(a) Close up of the chiral tweed texture generated by a monte carlo
simulation, and (b) its continuum representation by a mathematical
formula. On the shaded (clear) portions {\mdiv}\,{\bf m} is positive (negative) and
{\mcurl}\,{\bf m} is negative (positive). This shows the texture within a unit cell\,; the pattern repeats periodically to form the Tweed phase. }
\label{fig2 }
\end{center}
\end{figure}

\begin{figure}
\begin{center}
\includegraphics[width=3in]{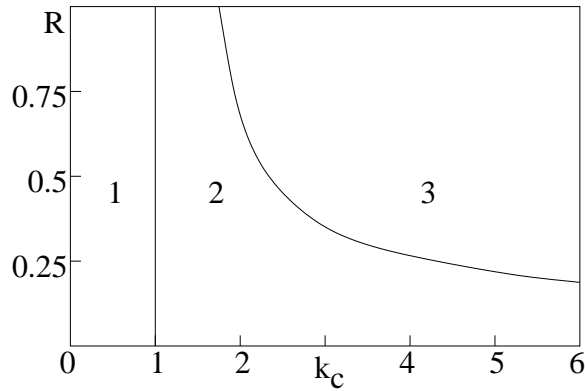}
\caption{Phases of the tilt texture domain with bulk chirality and $\sigma 
= 0$\,: (1) uniform phase (2) spiral defect
(where we have taken $\epsilon_{c} = 0$ and $r_{c} = 0.005$) (3) chiral tweed (stripe width $l^{*} = 0.01$).}
\label{texture-chiral}
\end{center}
\end{figure}
 
This texture may be easily parametrised variationally. First place the origin of coordinates on the boundary of the domain with the x-axis along a diameter,
and 
parametrise the texture by 
\begin{equation}
m_{x} = \vert \cos 
\frac{x}{l} \vert \,\, , \,\, m_{y} = -\vert \sin \frac{x}{l} \vert
\end{equation}
The texture consists of stripes of width $\pi l$ parallel to
the y-axis. This parametrisation is shown in Fig.\,7 for comparison with the optimal texture obtained in the simulation. This texture confers a 
net 
chiral strength to the domain --- 
$(\nabla\cdot\bf{m})(\nabla\times\bf{m})$ has the same sign everywhere, 
though the sign of individual terms, $\nabla\cdot\bf{m}$ and 
$\nabla\times\bf{m}$, varies from one stripe to the next.
The energy of 
the chiral tweed domain is 
\begin{equation} 
E = 2\pi R - \frac{k_{c}}{2l^2} 
\int_{0}^{2 R} dx \sqrt{2xR-x^2} \, \vert \sin \frac{2x}{l} \vert + \frac{\pi R^2}{2l^2}
\label{eq:energy2} 
\end{equation} 
The variational 
parameter $l$ approaches zero to minimise this energy --- higher order derivative terms in the hamiltonian, such as $\gamma\,(({\mdiv}\,{\bf{m}})^{4}+({\mcurl}\,{\bf{m}})^{4})$,
would however cutoff this monotonic decrease at some scale $l^{*}$. Note that the energy decreases rapidly as $k_c$ increases (Fig.\,9).
\begin{figure}
\begin{center}
\includegraphics[width=2in]{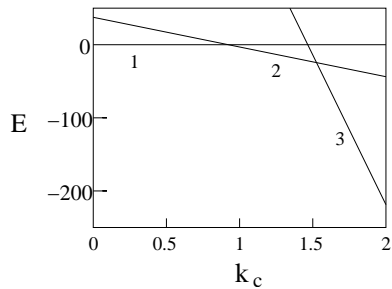}
\caption{Variation 
of the energy branches of the (1) Uniform (2) Spiral Hedgehog and (3) Tweed phases as a function of 
$k_{c}$ at $R = 1$ indicating discontinuous transitions. Notice that the 
energy decreases rapidly in phase (3) as $k_c$ increases.}
\label{fig2 }
\end{center}
\end{figure}

The phase diagram of this chiral domain is given in Fig.\,8,
together with a demonstration of discontinuous transitions
as seen by the crossing of the energy branches, Fig.\,9.

\section{Domain splitting}

So far our restriction to the strong segregation limit has also assumed that the planar domain of area $A$ does not break up. If the chiral strength is large enough, $k_c \gg 1$, the texture might prefer to maximise the number of spiral defect points. This could induce domain splitting. To study the conditions under which such breakup is favourable, we calculate the energy of $n$ circular domains of equal area, 
each bearing the
same spiral texture and compare it to a single circular domain
with the same total area and texture.
The total energy of this configuration is
\begin{equation}
E^{(n)} = 2 \pi \sigma_0 \sqrt{n} R - n \pi
\left(k_{c} - 1\right) \ln \frac{R/\sqrt{n}}{r_{c}} + n \epsilon_{c}
\end{equation}

For small values of $A$, a single domain $E^{(1)}$ has the least energy. As $A$ increases, $E^{(2)}$ becomes smaller than $E^{(1)}$ : chirality in the bulk
wins over interfacial energy (Fig.\,10).  As $A$ increases further, 
multidomain splitting is favoured.  At even higher values of
$A$, however, domain splitting is prohibited by a large interfacial energy cost; thus there is a window of areas for which chirality induces domain split up. This tendency to split
holds when $k_c$ is large enough; for a fixed value of $\sigma_0 r_c$, there is a critical $k_c$ beyond which chirality induced splitting would manifest.

This chirality induced splitting has implications in the
nucleation and growth of tilt-ordered domains in a fluid substrate, as in binary lipid mixtures where one of the components has a tilt and is chiral or in the nucleation of Sm-C$^*$ domains in a Sm-A film (since the SmA-SmC$^*$ transition can be first order \cite{smac}). For instance, consider a GUV composed of a binary lipid mixture of a saturated (minority component) and unsaturated lipid species, quenched below the gel transition of the saturated lipid. Domains of the gel phase spontaneously nucleate in the liquid-disordered phase. Growth initially occurs via coarsening; as the domains get larger, they undergo random brownian diffusion on the surface of the GUV, coalescing when two domains encounter each other.
If however the molecules in the gel phase have a tilt and chirality, then 
coalescence could be prevented by the above mechanism when the 
domains reach a size $R^*/\sqrt{2}$ (Fig.\,10). Further growth of the domain would be halted by this chirality induced domain repulsion. 
A similar feature should be 
observed in the nucleation and growth of Sm-C$^*$ domains in a Sm-A film.
\begin{figure}
\begin{center}
\includegraphics[width=3in]{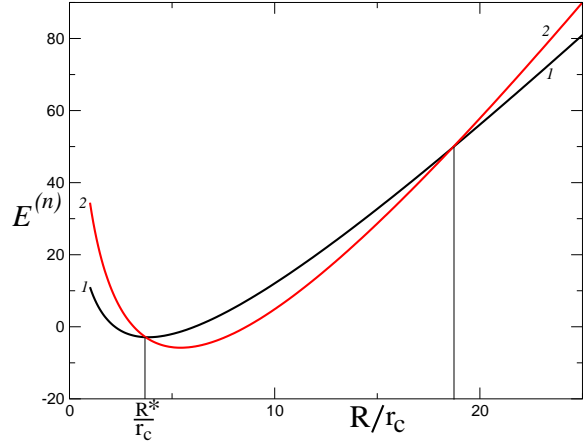}
\caption{Energy $E^{(n)}$, of a domain split into $n$
     equal parts, depends on its total area, $A = \pi R^{2}$. Here we show the 
    $n = 1$ and 2 branches : the domain prefers to split in two for a range of $R$ values ($\sigma_{0} r_{c} = 1.5$, $\epsilon_{c} = 5$, $k_{c} = 17$).}
\label{split}
\end{center}
\end{figure}

\section{Parameter Estimates}

In this section we estimate the various parameters entering the domain free-energy for typical systems under study.
We then compare the Frank energy scale to dipolar energies and provide some justification for why we have ignored
the latter. 

The values of the Frank constants entering the domain energy functional (\ref{eq:bulk}),  may be obtained from the corresponding values in bulk liquid
crystals. In a cholesteric liquid crystal, the director field
$\vec{n}$, describing the locally averaged molecular (long)-axis, describes a helical conformation about a fixed ordering axis. The energy density
of this conformation is given by the Frank expression \cite{degennes}
\begin{equation}
{\cal{E}} = \frac{K_{1}}{2} ({\mdiv}\,{\vec n})^{2} +
	    \frac{K_{2}}{2} ({\vec n} \cdot {\mcurl}\, {\vec n} + q)^{2} +
	    \frac{K_{3}}{2} ({\vec n} \times {\mcurl}\, {\vec n})^{2}
\end{equation}
where $2 \pi / q$ is the pitch of the helical conformation, much larger than the molecular size.

Comparing the energy density of the planar tilt-domain (\ref{eq:bulk}) with that of a cholesteric above, suggests the following correspondence : $k_{1} \sim a K_{1}$, and $k_{2} \sim a K_{3}$, where the length scale $a$ is of the order of
the thickness of the surface bearing the domain.
In addition, ${\cal{E}}$ has a term linear in $\mcurl\,{\vec n}$, suggesting that $k_{c} \sim a K_{2}$. Knowing the values of the Frank constants for the bulk cholesteric, $K_{1} \sim K_{2} \sim K_{3} \approx 10^{-11}$\,N, and taking
$a \approx 10 \AA$, we arrive at an estimate for the Frank coefficients in the energy functional (\ref{eq:bulk}),
$k_{1} \sim k_{2} \sim k_{c} \approx 10^{-20}$\,J. 
The magnitude of the chiral coefficient $k_c$ could be enhanced by the addition of chiral impurities.
Note that these values are 
almost an order of magnitude larger
than $k_{B}T$ at $30^{\circ}$\,C, and so  we may ignore the effect of
thermal fluctuations on the texture.

Estimates of the line tensions
may be obtained from studies of shape and texture changes 
 of tilt domains in Langmuir monolayers using Brewster angle microscopy \cite{LANG} and shapes of lipid domains in GUV's using fluorescence microscopy \cite{webb}. These methods give an estimate of $\sigma_{0} \approx 4 \times 10^{-13}$\,N for the isotropic line tension, and a value of $3.6 \times 10^4$\,m$^{-1}$ for the ratio $\sigma/K$. With our estimate of $10^{-20}$\,J for the Frank constant $K$, we arrive at an anisotropic line tension $\sigma \approx 3.6 \times 10^{-16}$\,N. 
 
With these parameter estimates, we now argue {\it post facto} that
long-ranged dipolar interactions are significantly weaker than the Frank contributions. 
Typical phospholipids have a dipole moment $\vert \vec{p} \vert \approx 1$\,debye 
\cite{tilt} directed roughly parallel to the plane of the membrane.
The strength of dipolar interactions of neighbouring tilted phospholipids is of the order $\vert \vec{p} \vert ^{2}/a^{3}$ where $a \approx 10 \AA$ is the separation of the lipid dipoles. This energy is of the order of $10^{-22}$\,J, as a result, dipolar interactions can not challenge the order imposed on ${\bf m}$ by the Frank energy. However if the molecules are charged, then the strength of electrical interaction may be considerably enhanced $\vert \vec{p} \vert q/a^{2} \approx 10^{-20}$\,J for $q = 1$\,coulomb and $a \approx 8 \AA$, giving rise to significant charge-dipole or dipole-dipole interactions. Colloidal systems with substantial dipolar effects are known to exhibit string-like aggregates \cite{yethiraj}. 

\section{Discussion}

In this paper we have discussed the equilibrium shapes and textures of a single domain consisting of molecules described by a tilt embedded on a flat substrate containing non-tilt molecules. In addition to having a tilt, the molecules constituting the domain may be chiral. The interplay between tilt, geometry and chirality give rise to a rich variety of shapes and textures separated by discontinous transitions. We have arrived at our results by a combination of Monte Carlo simulations and {\it
exact} variational calculations. In addition our variational ansatz allows for multiply connected domains.
Our key results : (i) In the achiral limit, optimum textures include virtual boojum, annular and hedgehog phases. A novel feature is the occurence of spontaneous chiral symmetry breaking to  a chiral annular phase. (ii) When the 
domain shapes are allowed to deviate from circularity, then the tilt induced anisotropic line tension typically give rise to domain shapes which are flattened and elongated. (iii) Chirality produces spiral defects and an unusual chiral tweed phase. (iv) Chirality induces large enough domains to breakup into smaller domains leading to a limiting domain size (keeping all other parameters fixed) and preventing coalescence.

On symmetry grounds, one may argue \cite{meyer,degennes} that the tilt phase of chiral molecules having a permanent dipole moment is {\it ferroelectric}. Thus an externally applied electric field ${\vec E}$ can couple to the polarization vector
${\vec P}$, via an ${\vec E}\cdot {\vec P}$ coupling.
The polarization vector ${\vec P}$ is perpendicular to both the tilt ${\vec m}$ and the normal to the plane ${\hat z}$ \cite{degennes}, i.e., ${\vec P} = \alpha {\hat z} \times {\vec m} \equiv \alpha (\sin \phi, \cos \phi, 0)$. The total energy (\ref{eq:bulk}), (\ref{eq:interface}) must be augmented by $- \alpha (E_x \sin \phi + E_y \cos \phi)$, in addition to an electrostatic self energy proportional to $P^2$. In the low chirality limit $k_c < 1$, this field imposes a preferred orientation for ${\bf m}$; by changing the direction of the field, one may switch between
the two textures \cite{electric}.
In addition, strongly chiral systems admit a pseudo-scalar coupling of the form $({\vec m} \cdot  \nabla {\vec m} \cdot  {\vec P})$, which is simply $(\nabla \times {\vec m})_z$ \cite{degennes}; this leads to a switch in the {\it handedness} of the
spiral texture on changing the direction of the external field.

While the results obtained here are applicable to a variety of systems featuring tilt domains of finite size embedded in a 2d substrate, our main interest is in the shapes and dynamics of domains in a multicomponent lipid mixture. Giant unilamellar vesicles composed of saturated and unsaturated lipid components, typically exhibit a wide coexistence regime between a liquid disordered
phase (enriched in the unsaturated component) and a gel phase (enriched in the saturated lipid). Following a quench into this coexistence region, domains of the minority component nucleate and grow in the majority phase. If the saturated lipid is chiral and has a tilt in the gel phase, then we believe that the analysis carried out here has some relevance to the shapes and dynamics of nucleating domains \cite{sanat,vernita}.

\section{Acknowledgements} 

We thank Y. Hatwalne, N.V. Madhusudhana, V.A. Raghunathan and S. Ramaswamy for useful discussions. 
We thank V. Gordon, S. Egalhaaf and W. Poon, and V.A. Raghunathan and S. Karmakar, for sharing their experimental results prior to publication.

\end{document}